\documentclass[twocolumn,twocolappendix,times]{aastex631}

\usepackage{graphicx}	% Including figure files
\usepackage{amsmath}	% Advanced maths commands
\usepackage{amssymb}	% Extra maths symbols
\usepackage{xcolor}
\usepackage{ctable}
\usepackage{booktabs}
\usepackage{threeparttable}
\usepackage{comment}
\usepackage{newtxtext,newtxmath}

\begin{document}

\title{X-raying CAMELS: Constraining Baryonic Feedback in the Circum-Galactic Medium 
\\ with the CAMELS simulations and eRASS X-ray Observations}

\correspondingauthor{Erwin T.\ Lau}
\email{erwin.lau@cfa.harvard.edu}

\author[0000-0001-8914-8885]{Erwin T.\ Lau}
\affiliation{Center for Astrophysics, $\vert$ Harvard \& Smithsonian, 60 Garden Street, Cambridge, MA, 02138, USA}
\affiliation{University of Miami, Department of Physics, Coral Gables, FL 33124, USA}

\author[0000-0002-6766-5942]{Daisuke Nagai}
\affiliation{Department of Physics, Yale University, New Haven, CT 06520, USA}
\affiliation{Department of Astronomy, Yale University, New Haven, CT 06520, USA}

\author[0000-0003-0573-7733]{\'Akos Bogd\'an}
\affiliation{Center for Astrophysics, $\vert$ Harvard \& Smithsonian, 60 Garden Street, Cambridge, MA, 02138, USA}

\author{Isabel Medlock}
\affiliation{Department of Astronomy, Yale University, New Haven, CT 06520, USA}

\author[0000-0002-3391-2116]{Benjamin D. Oppenheimer}
\affiliation{Center for Astrophysics and Space Astronomy, 
University of Colorado, 389 UCB,
Boulder, CO, 80309, USA}

\author[0000-0001-5846-0411]{Nicholas Battaglia}
\affiliation{Department of Astronomy, Cornell University, 404 Space Sciences Building, Ithaca, NY 14853, USA}

\author[0000-0001-5769-4945]{Daniel Angl\'es-Alc\'azar}
\affiliation{Department of Physics, University of Connecticut, 196 Auditorium Road, U-3046, Storrs, CT, 06269, USA}

\author[0000-0002-3185-1540]{Shy Genel}
\affiliation{Center for Computational Astrophysics, Flatiron Institute, 162 Fifth Ave, New York, NY, 10010, USA}
\affiliation{Columbia Astrophysics Laboratory, Columbia University, 550 West 120th Street, New York, NY 10027, USA}

\author[0000-0001-7899-7195]{Yueying Ni}
\affiliation{Center for Astrophysics, $\vert$ Harvard \& Smithsonian, 60 Garden Street, Cambridge, MA, 02138, USA}

\author[0000-0002-4816-0455]{Francisco Villaescusa-Navarro}
\affiliation{Center for Computational Astrophysics, Flatiron Institute, 162 Fifth Ave, New York, NY, 10010, USA}

% Abstract of the paper
\begin{abstract}
The circumgalactic medium (CGM) around massive galaxies plays a crucial role in regulating star formation and feedback. Using the CAMELS simulation suite, we develop emulators for the X-ray surface brightness profile and the X-ray luminosity--stellar mass scaling relation to investigate how stellar and AGN feedback shape the X-ray properties of the hot CGM. Our analysis shows that at CGM scales ($10^{12} \lesssim M_{\rm halo}/M_\odot \lesssim 10^{13}$, $10\lesssim r/{\rm kpc} \lesssim 400$), stellar feedback more significantly impacts the X-ray properties than AGN feedback within the parameters studied.
Comparing the emulators to recent eROSITA All-Sky Survey observations, it was found that stronger feedback than currently implemented in the IllustrisTNG, SIMBA, and Astrid simulations is required to match observed CGM properties. However, adopting these enhanced feedback parameters causes deviations in the stellar-mass-halo-mass relations from observational constraints below the group mass scale. This tension suggests possible unaccounted systematics in X-ray CGM observations or inadequacies in the feedback models of cosmological simulations. 
\end{abstract}


%%%%%%%%%%%%%%%%%%%%%%%%%%%%%%%%%%%%%%%%%%%%%%%%%%

%%%%%%%%%%%%%%%%% BODY OF PAPER %%%%%%%%%%%%%%%%%%

\section{Introduction}

The circumgalactic medium (CGM) around massive galaxies plays a fundamental role in galaxy evolution \citep[][for reviews]{tumlinson_etal17,faucher-giguere_oh}. It serves as an important reservoir for star formation and a repository for metal-enriched outflows driven by feedback from supernova (SN) and active galactic nuclei (AGN). In particular, the hot phase of the CGM can serve as a unique probe of baryon cycles in massive galaxies \citep[e.g.,][for a review]{donahue_voit22}. Constraining and understanding the physical properties of the CGM is thus crucial for understanding galaxy evolution and structure formation in the Universe. 

Multi-wavelength observations have revolutionized our understanding of CGM physical properties, particularly through X-ray observations \citep[e.g.,][]{bogdan_etal13b,bogdan_etal13a,anderson_etal15,bogdan_etal17} and the Sunyaev-Zeldovich (SZ) effects \citep[e.g.,][]{amodeo_etal21,bregman_etal22,das_etal23}.
Despite its relatively low angular resolution compared to {\em Chandra} and XMM-{\rm Newton}, X-ray observations with the eROSITA/SRG telescope have advanced the state of CGM observations in X-ray through its large sky coverage. Recent stacked X-ray observations from the eROSITA Final Exposure Depth Survey (eFEDS) fields \citep{chadayammuri_etal22,comparat_etal22} inferred that the CGM X-ray luminosity seems to be lower than that of cosmological simulations at higher stellar mass halos. However, these results are still uncertain given the statistical uncertainties due to cosmic variances in the stacking procedures. The recently released stacked X-ray surface brightness profiles and the X-ray luminosities from the half-sky eROSITA All Sky Survey (eRASS) from \citet{eRASS_CGM1, eRASS_CGM2} represent state-of-the-art X-ray observations of the CGM to date. Compared to the 140 square degree eFEDS field, the half-sky coverage of eRASS contains a much larger galaxy sample for stacking, which increases the signal-to-noise ratio by a factor of a few, reducing the statistical uncertainty on the stacked signal, thus potentially allowing for better constraints on CGM properties and their physics.
 
The CAMELS (Cosmology and Astrophysics with MachinE Learning Simulations) suite \citep{camels_multifields, camels, camels_astrid}, which consists of a large number of cosmological hydrodynamical simulations with varying cosmological parameters and baryonic feedback, is uniquely suited for studying feedback physics at the CGM scale. 
The CAMELS suite has been used to study the impact of feedback on CGM properties in SZ \citep{moser_etal22,pandey_etal23,wadekar_etal23} and in radio through Fast Radio Bursts \citep[FRB:][]{medlock24a}. CAMELS also enables the development of machine learning techniques for predicting CGM properties with X-ray and HI maps \citep{gluck_etal23}, characterizing the impact of feedback on matter power clustering \citep{delgado_etal24,gebhardt_etal24}, and breaking the degeneracies between feedback and cosmological parameters via kinetic SZ and FRB \citep{nicola_etal22}. Zoomed-in simulations from CAMELS \citep{lee_etal24} also predict how X-ray and SZ profiles of clusters and groups can be used to constrain cosmology and feedback physics \citep{hernadez-martinez_etal24}.  

In this paper, we explore the dependence of CGM X-ray properties on feedback models using CAMELS, and use the latest eRASS observations to constrain the CAMELS feedback parameters. We study how different subgrid models of feedback impact the X-ray CGM emission, and assess how the latest stacked X-ray surface brightness (XSB) and X-ray luminosities from eRASS are able to constrain CGM feedback physics in CAMELS. To do so, we follow the approach of \citet{moser_etal22} by constructing emulators of the XSB profiles and X-ray luminosity from CAMELS as a function of the feedback parameters. By comparing the X-ray emulated observables to the eRASS data, we derive constraints on the feedback parameters. 

\section{Methodology}\label{sec:methods}

\subsection{Simulations}\label{sec:sims}

The Cosmology and Astrophysics with MachinE Learning Simulations (CAMELS) dataset \citep{camels_multifields, camels, camels_astrid} consists of cosmological simulations with different $\Lambda$CDM cosmologies, hydro-solvers, and subgrid feedback physics. Each CAMELS simulation we use follows the evolution of a $(25 h^{-1}\mathrm{\,comoving\,Mpc})^3$ volume with base parameters $\Omega_b = 0.049$, $h = 0.6711$, and $n_s = 0.9624$ assuming a flat $\Lambda$CDM universe. The other two main parameters $\Omega_M$ and $\sigma_8$ are varied in the simulations. 
All simulations follow the evolution of $256^3$ dark matter particles and fluid elements, corresponding to DM particle resolution of $M_{\rm DM} = 6.49\times 10^7 (\Omega_M-\Omega_b)/0.251\,h^{-1}M_\odot$ and gas mass resolution of $1.25\times 10^{7}h^{-1}M_\odot$.

\subsubsection{Simulation Suites}\label{sec:suites}

These simulations were run using different simulation codes. For this work, we focus on IllustrisTNG \citep{pillepich_etal18, nelson_etal18}, SIMBA \citep{dave19}, and Astrid \citep{bird_etal22,astrid}, which had been run using the AREPO, GIZMO, and MP-Gadget codes, respectively.  We used the Latin Hypercube (LH) sets for each of the three runs, each consists of 1000 simulations with different parameters sampled using Latin Hypercube. These parameters include $\Omega_m$ and $\sigma_8$, two parameters corresponding to supernovae feedback and another two to AGN feedback. There are also fundamental differences between the different simulation suites. In IllustrisTNG, the hydrodynamic equations are solved on a moving mesh; in SIMBA they are solved using meshless methods, and in Astrid they are solved using the pressure-entropy formulation of smoothed particle hydrodynamics (pSPH). Additionally, each simulation suite has different subgrid physics; for example, SIMBA tracks dust grains, while IllustrisTNG includes magnetohydrodynamics. 

\begin{table*}
\centering
\renewcommand{\arraystretch}{1.5} % Adjust row spacing
\setlength{\tabcolsep}{10pt}     % Adjust column spacing
\begin{tabular}{| c | p{4.5cm} | p{4.5cm} | p{4.5cm} |}
%\begin{tabular}{| c | c | c | c |}
\hline
\textbf{Parameter} &
\textbf{IllustrisTNG} &
\textbf{SIMBA} &
\textbf{Astrid} \\
\hline
\textbf{SN1} &
Galactic winds: energy per unit SFR [0.25, 4.0] &
Galactic winds: mass loading [0.25, 4.0] &
Galactic winds: energy per unit SFR [0.25, 4.0] \\
%\hline
\textbf{SN2} &
Galactic winds: wind speed [0.5, 2.0] &
Galactic winds: wind speed [0.5, 2.0] &
Galactic winds: wind speed [0.5, 2.0]\\
%\hline
\textbf{AGN1} &
Kinetic mode BH feedback: energy per unit BH accretion [0.25, 4.0] &
QSO \& jet mode BH feedback: momentum flux [0.25, 4.0] &
Kinetic mode BH feedback: energy per unit BH accretion [0.25, 4.0] \\
%\hline
\textbf{AGN2} &
Kinetic mode BH feedback: ejection speed / burstiness [0.5, 2.0] &
Jet mode BH feedback: jet speed  [0.5, 2.0] &
Thermal mode BH feedback: energy per unit BH accretion [0.25, 4.0] \\
\hline
\end{tabular}
\caption{Parameters associated with different simulations and feedback mechanisms.}
\label{tab:simulation_parameters}
\end{table*}

We focus on the four key feedback parameters explored in the CAMELS suites: two parameters for stellar feedback: SN1, SN2, and two parameters for AGN feedback: AGN1, AGN2.  
Here, we give a qualitative overview of the differences between the feedback implementation between the different simulation suites. We refer the reader to \citet{camels_astrid,medlock24b} for more detailed comparison of the models. 
Here we provide a qualitative overview of how the three simulation suites implement their stellar and black hole feedback, and highlight the most salient differences between them. 
To avoid confusion between the different feedback physics in the three simulation suites, hereafter, we denote the feedback parameters of the CAMELS-IllustrisTNG run as $A_{\rm SN1}$, $A_{\rm SN2}$, $A_{\rm AGN1}$, $A_{\rm AGN2}$, for the CAMELS-SIMBA run as $B_{\rm SN1}$, $B_{\rm SN2}$, $B_{\rm AGN1}$, $B_{\rm AGN2}$, for the CAMELS-Astrid run as $C_{\rm SN1}$, $C_{\rm SN2}$, $C_{\rm AGN1}$, $C_{\rm AGN2}$. For all parameters, a value of $1$ represents the fiducial value that was adopted in their original simulations. 

\subsubsection{Stellar Feedback Models}\label{sec:sn_feedback}

All three suites implement stellar feedback as galactic winds. For CAMELS-IllustrisTNG, the $A_{\rm SN1}$ parameter represents a prefactor that controls the energy per unit star formation rate (SFR), for CAMELS-SIMBA, the $B_{\rm SN1}$ represents the wind mass outflow rate per unit SFR, i.e. the mass-loading factor; for CAMELS-Astrid, the $C_{\rm SN1}$ controls the energy per unit SFR, same as CAMELS-IllustrisTNG. In all three suites, the SN2 parameter represents the normalization factor for the galactic wind speed. Consequently, the stellar feedback model is essentially the same for CAMELS-IllustrisTNG and CAMELS-Astrid, and slightly different for CAMELS-SIMBA. Specifically, keeping SN1 constant while varying the SN2 parameter maintains a fixed wind energy output but changing the wind speed for both CAMELS-IllustrisTNG and CAMELS-Astrid, but for CAMELS-SIMBA, while the mass loading factor remains constant for a fixed SN1, changing SN2 varies both the wind speed and the wind energy. 

\subsubsection{AGN Feedback Models}\label{sec:agb_feedback}

The AGN feedback models differ substantially between the three simulation suites. In CAMELS-IllustrisTNG, the AGN feedback is implemented as a kinetic feedback from supermassive black holes (SMBHs) seeded onto massive halos. The SMBHs in IllustrisTNG accrete assuming a spherical Bondi accretion model. The AGN feedback mode switches from thermal to kinetic modes from high to low SMBH accretion rate. The $A_{\mathrm{AGN1}}$ parameter controls the energy output of the AGN feedback per unit SMBH accretion in the kinetic mode. Kinetic feedback is implemented by injecting kinetic energy around the SMBH in a random direction. The kinetic energy is injected until a minimum amount of energy is accumulated to ensure that the energy injection events are powerful and independent of each other. This minimum amount of energy is controlled by the $A_{\mathrm{AGN2}}$ parameter. Thus, this $A_{\mathrm{AGN2}}$ can be thought of as controlling the ``burstiness'' in the kinetic feedback of AGN in IllustrisTNG. 

On the other hand, CAMELS-SIMBA implements AGN feedback differently from IllustrisTNG. The SMBH in SIMBA has two accretion modes: Bondi accretion and gravitational torque accretion. The kinetic feedback mode in CAMELS-SIMBA switches from wind/QSO mode to jet mode from high to low SMBH accretion rate. The $B_{\mathrm{AGN1}}$ parameters control the momentum flux of feedback per unit SMBH accretion for both jet and wind modes, and the $B_{\mathrm{AGN2}}$ controls the speed of the jet. 

For CAMELS-Astrid, the accretion rate onto the SMBH is estimated via Bondi accretion. The SMBH feedback also follows a two-mode approach that switches from thermal to kinetic feedback modes from high to low SMBH accretion rate. The $C_{\mathrm{AGN1}}$ and $C_{\mathrm{AGN2}}$ modulate the efficiency of kinetic and thermal feedback separately. The $C_{\mathrm{AGN1}}$ has the same physical meaning as $A_{\mathrm{AGN1}}$ in CAMELS-IllustrisTNG. Compared to CAMELS-IllustrisTNG, CAMELS-Astrid turns on the AGN kinetic at a lower SMBH accretion rate threshold with a lower upper limit of feedback efficiency, resulting in a milder AGN kinetic feedback compared to CAMELS-IllustrisTNG. 

In Table~\ref{tab:simulation_parameters}, we summarize the physical meaning of the four feedback parameters in CAMELS-IllustrisTNG, CAMELS-SIMBA, and CAMELS-Astrid, and the ranges of the values adopted in the CAMELS runs. 

\subsubsection{Cosmological Parameters}\label{sec:cosmo_params}

Finally, for the two cosmological parameters, $\Omega_M$ varies between $0.1$ and $0.5$, and $\sigma_8$ varies between $0.6$ and $1.0$. The fiducial value for the feedback parameters is $1.0$, and $0.3$ and $0.8$ for $\Omega_M$ and $\sigma_8$ respectively. 

\subsection{Modeling X-ray Observables}\label{sec:xray_obs}

The X-ray Surface Brightness (XSB) profile as a function of projected radius $R_p = [10,10^3]$~kpc, binned in 20 uniform logarithmic bins, for each halo is computed from the simulation as 
\begin{equation}
    S_{X}(R_p) = 2\int_{R_p}^{\infty} \frac{\epsilon(r)r}{\sqrt{r^2-R_p^2}} dr,
\end{equation}
where
\begin{equation}
    \epsilon(r) = 4\pi \int n_e(r) n_H(r) r^2 \Lambda(T(r), Z(r))dr ,
\end{equation}
is the 3D X-ray emissivity of the CGM as a function of the distance from the halo center $r$; $n_e$ and $n_H$ are the electron and hydrogen densities, respectively; $\Lambda$ is the X-ray cooling function, which is a function of CGM temperature $T$ and metallicity $Z$, computed using the APEC model \citep{foster_etal12}. For each halo, we compute the densities, temperatures, and metallicity profiles (assuming spherical symmetry) for the hot phase gas with temperature $>10^5$~K. The energy range of the emission in the observer's frame is chosen to be $ E = [0.5, 2.0]$~keV. The X-ray profiles are computed using the {\tt CGM Toolkit} code \citep{lau_cgmtoolkit}. 
Note that at the halo mass scales in which we are interested ($\log_{10}M/M_\odot< 13.5$) metal lines dominate the X-ray emission \citep[see][]{lovisari_etal21}. 

We compute the X-ray luminosity by integrating the XSB profile up to $R_{500c}$:
\begin{equation}
    L_X(<R_{500c}) = \int_0^{R_{500c}} S_{X}(R_p) 2\pi R_p dR_p.
\end{equation}

\subsection{Emulation}\label{sec:emulation}

We follow the same approach in constructing a Gaussian process emulator for the XSB profile as in \citet{moser_etal22}, where an emulator for SZ profiles was developed to capture the dependence of feedback physics. The emulator can be thought of as a multidimensional interpolator of the XSB profile. 

\subsubsection{XSB Profiles}

We construct emulators for the XSB profiles over a total of 8 parameters: the 4 feedback parameters, the 2 cosmological parameters: $\Omega_M$ and $\sigma_8$, and 2 halo parameters: redshift $z$ and stellar mass of the central subhalo $M_{\star}$. For each of the 1000 LH simulations with different cosmological and feedback parameters, we computed the mean XSB profiles over the 3 available snapshots corresponding to $z = [0.10, 0.15, 0.18]$ that matches the redshift range of the eRASS CGM data, with 6 uniformly spaced stellar mass bins spanning the range $\log_{10}(M_{\star}/M_{\odot}) \in [9.0,12.0]$, covering the range of stellar masses in the eRASS CGM observation $\log_{10}(M_{\star}/M_{\odot}) \in [10.5,11.5]$. The resulting stellar mass bins have a large enough number of halos per bin to have stable trends in the profiles -- changing the minimum and maximum stellar mass in the binning scheme to $\log_{10}(M_{\star}/M_{\odot}) \in [10.5,11.5]$ does not change the prediction of the emulator. The small box size ($25\,h^{-1}{\rm Mpc}$) of CAMELS limits our mass selections to objects with halo masses $M_{200c}\lesssim 10^{13} M_{\odot}$, and the higher-mass bins have significantly fewer objects than the lower-mass bins. The number of halos in the bins ranges from $30$ to $843$, depending on the specific run and snapshot. This results in 9000 (1000 runs times 3 redshifts times 3 mass bins) mean XSB profiles as input to the emulation. Following the SZ emulator of \citet{moser_etal22}, each input XSB profile is represented in basis vectors, and the weights of the basis vectors are decomposed using principal component analysis (PCA). The weights are interpolated using a radial basis function (RBF) interpolator. Principal components are sorted by decreasing variance, and the number of PCA components is a free parameter, which impacts the amplitude of residual variance. We used 12 PCA components and also checked that adding more components does not affect the emulator results. The PCA components are then emulated over the parameter space of the LH set using Gaussian Process. 

\subsubsection{X-ray Luminosity-Stellar Mass Relation}

We also emulate the X-ray Luminosity-Stellar Mass $L_X-M_\star$ relation. Similarly to the XSB profile emulation, we compute the $L_X$ values for halos in the LH simulation for the same redshift snapshots. The difference from the XSB emulation is that we emulate $L_X$ in 8 logarithmic stellar mass $\log_{10}(M_{\star}/M_{\odot}) \in [9,12]$ bins, compare to the 6 stellar mass bins in XSB. This enables us to better capture the local changes in slope of the scaling relation.  We perform emulation in the same way as in the XSB profile: we decomposed each relation into 12 PCA components, which are then emulated over the parameter space using the same Gaussian Process as the XSB profiles.  We have checked that the $L_X$ and XSB emulator are consistent with each other as we can recover the same $L_X$ values by integrating the XSB emulator emulated profile for a given stellar mass.

Note that given the relatively small box size, for some LH runs, there are very few halos in the highest stellar mass bin, especially when the feedback values are extremely high or low. This could result in zero values of the emulated XSB profiles or $L_X$ that breaks the emulator. To circumvent this, we add a small number ($10^{-30}$) to the zero values. We have checked that this does not affect the accuracy of the emulator.  

The {\tt CAMELS CGM Emulator} code \citep{lau_camels_cgm_emulator} used in this work is publicly available on GitHub\footnote{\url{https://github.com/ethlau/CAMELS_emulator}}.

\section{Observation Data from eRASS}\label{sec:eRASS_obs}

The eROSITA survey with its half-sky coverage overlaps with large-scale galaxy surveys such as the Sloan Digital Sky Survey (SDSS). This uniquely enables the stacking of X-ray emissions on a large number ($\sim 10^5$) of optically identified galaxies, resulting in high signal-to-noise measurements of the X-ray CGM. 

We use the `CEN' sample from the eRASS CGM papers \citep{eRASS_CGM1, eRASS_CGM2} for comparison. The sample is based on the spectroscopic galaxy sample of SDSS MGS (Main Galaxy Sample). The sample for the stacked XSB profiles consists of three main bins based on stellar mass: `MW', `M31', and `2M31', corresponding to the median stellar mass of $\log_{10} (M_{\star, {\rm med}}/M_\odot) = 10.74, 11.11, 11.34$, with median redshifts $z_{\rm med} = 0.08, 0.12, 0.15$, and the number of stacked galaxies = $30825, 26099, 20342$, respectively. 
The X-ray luminosity-stellar mass scaling relation data includes an additional lower stellar mass bin with $\log_{10} (M_{\star, {\rm med}}/M_\odot) = 10.26 $ with $z_{\rm med} = 0.05$ and 7956 galaxies. 
We refer the reader to \cite{eRASS_CGM1, eRASS_CGM2} for details on data reduction, sample selection, and discussion of the systematics of observation. 

\section{Results}\label{sec:results}

\begin{figure*}
    \centering
    \includegraphics[width=1.0\textwidth]{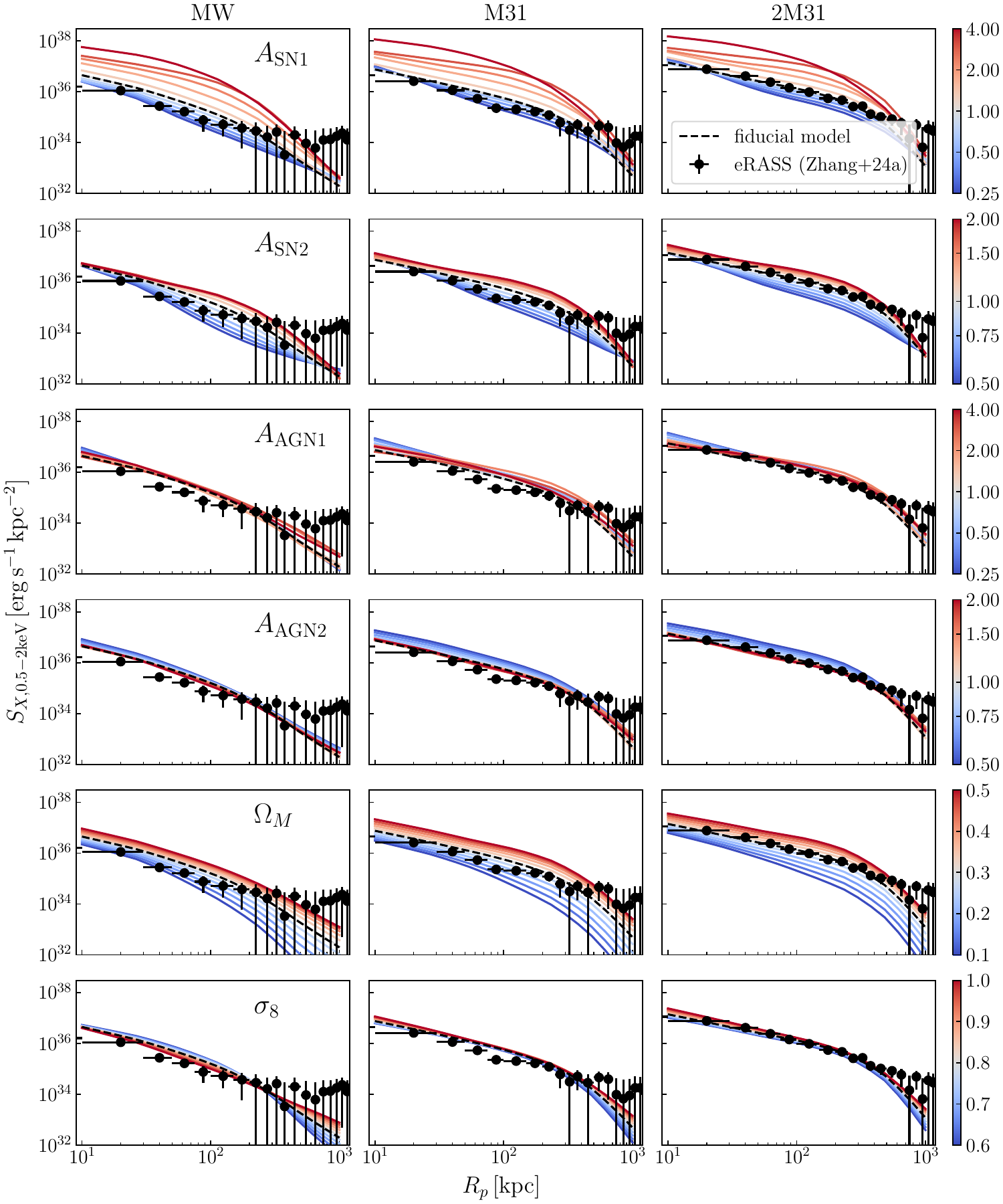}
    \caption{Comparison of the X-ray surface brightness (XSB) profiles from the eRASS CGM XSB profiles in black data points with 1$\sigma$ errorbars \citep{eRASS_CGM1} with the emulated profiles for the CAMELS-IllustrisTNG with varying values for the 4 feedback parameters: $A_{\rm SN1}$, $A_{\rm SN2}$, $A_{\rm AGN1}$, and $A_{\rm AGN2}$, showed in panels from top 1st to 4th rows. The colors indicate the strengths of the feedback parameters, which vary from $[0.25, 4.0]$ for $A_{\rm SN1}$, $A_{\rm AGN1}$, and $[0.5, 2.0]$ for $A_{\rm SN2}$, and $A_{\rm AGN2}$. The bottom two rows show the dependence on $\Omega_M$ and $\sigma_8$. The black dashed lines indicate where the parameters adopt their fiducial values. The different columns show the comparisons in the three different stellar mass bins: `MW', `M31', and `2M31', for the left, middle, and right panels, respectively. }
    \label{fig:tng_xsb_profiles}
\end{figure*}

\begin{figure*}
    \centering
    \includegraphics[width=1.0\textwidth]{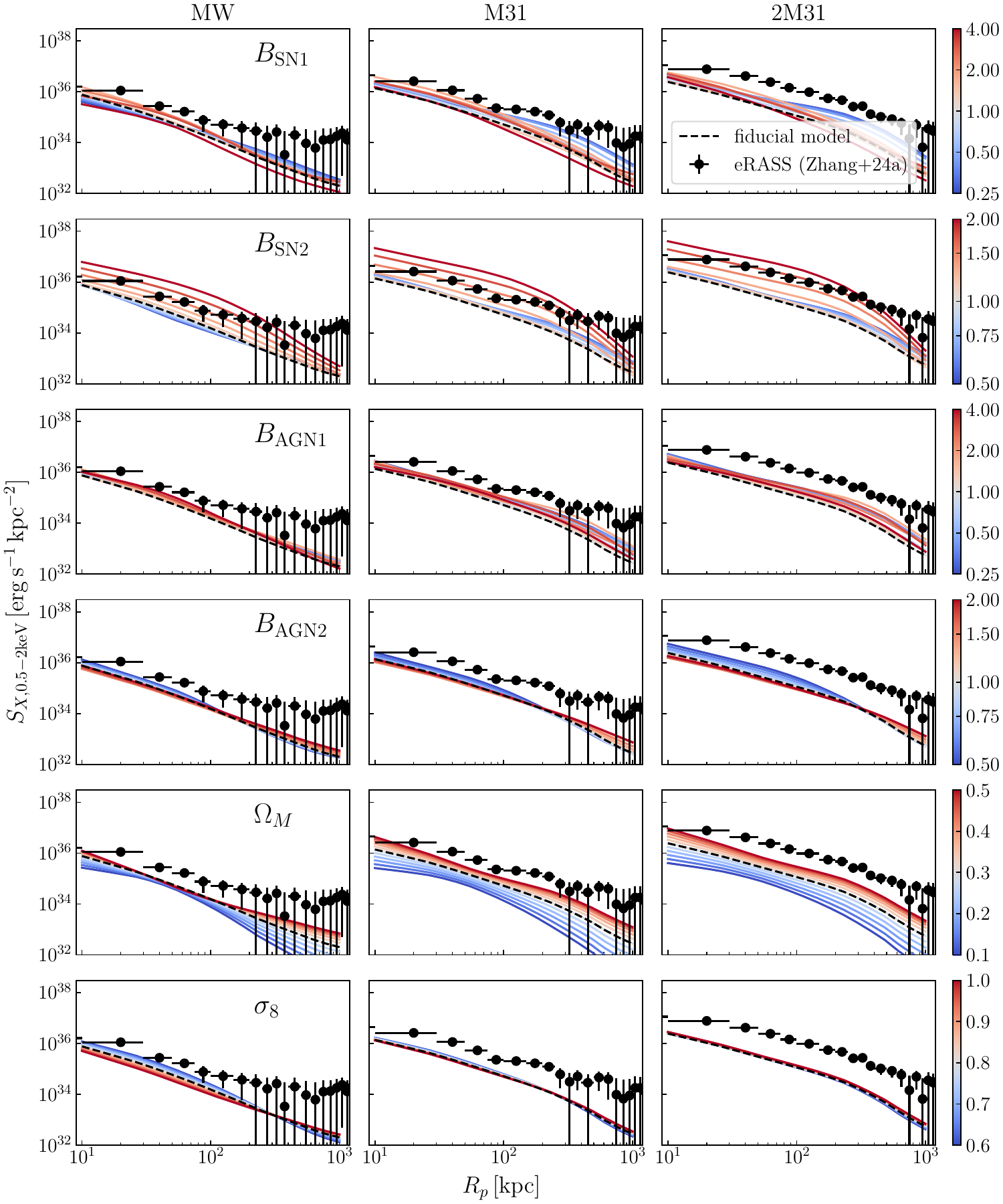}
    \caption{
    Same as Figure~\ref{fig:tng_xsb_profiles} but for CAMELS-SIMBA.
    }
    \label{fig:simba_xsb_profiles}
\end{figure*}

\begin{figure*}
    \centering
    \includegraphics[width=1.0\textwidth]{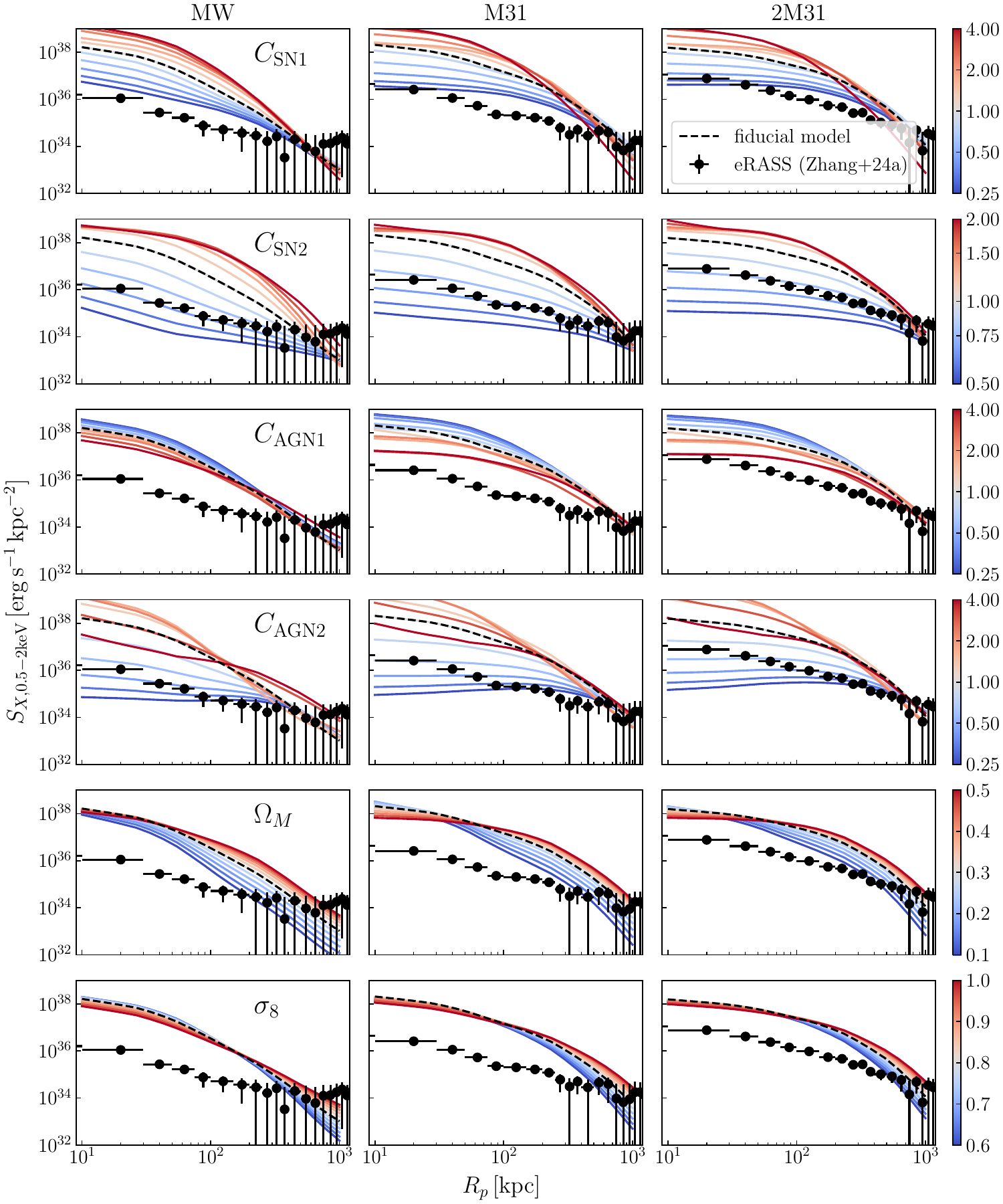}
    \caption{
    Same as Figure~\ref{fig:tng_xsb_profiles} but for CAMELS-Astrid.    
    }
    \label{fig:astrid_xsb_profiles}
\end{figure*}

\subsection{XSB Profiles}

In this section, we demonstrate how the XSB profiles from CAMELS-IllustrisTNG, CAMELS-SIMBA, and CAMELS-Astrid depend on the feedback parameters and how they compare with the eRASS observation. 

\subsubsection{CAMELS-IllustrisTNG}\label{sec:tng}

The top row in Figure~\ref{fig:tng_xsb_profiles} shows the comparisons of the eRASS CGM profiles to the emulated XSB profiles based on the CAMELS-IllustrisTNG runs in the three stellar mass bins. Among all feedback parameters, the XSB profile is most dependent on $A_{\rm SN1}$, which represents the supernova energy imparted to the surrounding gas per unit star formation rate (SFR). Increasing $A_{\rm SN1}$ suppresses the AGN feedback by reducing the gas available for accretion onto the supermassive black holes (SMBHs) in the halo centers, as shown in \citet{lee_etal24, medlock24b}, thus increasing the overall gas density. We have checked the CGM temperature also increases with $A_{\rm SN1}$, but not as much as gas density. As the XSB depends on the density squared, and less so on temperature, the increase in gas density leads to higher XSB.  In terms of mass dependence, the XSB profile in the least massive `MW' bin is more sensitive to variation in $A_{\rm SN1}$ compared to the other two stellar mass bins, due to the shallower gravitational potential of MW-size halos. 

The second row shows the comparisons of the eRASS CGM XSB profiles with the emulated XSB profiles with varying $A_{\rm SN2}$ feedback, which represents the speed of the galactic wind. Increasing $A_{\rm SN2}$ leads to an overall increase in the XSB, particularly in the radial range $[10,300]$~kpc, for all stellar mass bins. Increasing the SN wind speed also suppresses the AGN feedback by decreasing the available gas for SMBH accretion, but not enough to drive the hot gas out of the halo potential; therefore, the hot gas density, thus the XSB, increases with increasing SN wind speed. In addition to the amplitude of the XSB profile, the shape of the eRASS XSB profile matches well with the emulated profiles with higher $A_{\rm SN2}$, which shows flatter profiles in $[10,300]$~kpc. The impact of $A_{\rm SN2}$ is more prevalent in the lowest stellar mass `MW' bin with shallower gravitational potential.  

The third row shows the comparisons of the eRASS CGM profiles with the emulated XSB profiles with varying $A_{\rm AGN1}$ feedback in CAMELS-IllustrisTNG. The $A_{\rm AGN1}$ represents the power of the AGN feedback in the kinetic mode. The fiducial value $A_{\rm AGN1}=1$ (while other feedback parameters are kept fixed at their respective fiducial values) is able to match the observed eRASS profiles in the MW stellar mass bin, but slightly overpredicts the profile in the other two higher stellar mass bins.  Varying $A_{\rm AGN1}$ has relatively little impact on the XSB profile across all mass bins, with the least massive `MW' bins being affected the least. For the highest mass bin `2M31', increasing $A_{\rm AGN1}$ changes the shape of the XSB profile, making the profile flatter outside the core. 

The fourth row shows the comparisons of the eRASS CGM profiles to the emulated XSB profiles with varying $A_{\rm AGN2}$ feedback, which represents the `burstiness' of the AGN feedback. Specifically, a higher $A_{\rm AGN2}$ means less frequent but more powerful AGN feedback events. The higher $A_{\rm AGN2}$ leads to a decrease in amplitude in the emulated XSB profiles at all stellar mass bins, as the less frequent but more powerful AGN feedback are more capable in pushing the gas out of the potential well, reducing the gas density and thus the X-ray emission. The impact of $A_{\rm AGN2}$ is more prevalent in the highest stellar mass bin `2M31'. 

\subsubsection{CAMELS-SIMBA}\label{sec:simba}

Figure~\ref{fig:simba_xsb_profiles} shows the comparisons of the eRASS CGM profiles to the emulated XSB profiles based on the CAMELS-SIMBA runs. 

The top row shows the XSB dependence on $B_{\rm SN1}$ the energy of the supernova feedback per unit stellar mass, i.e.\ the mass loading factor. Similarly to CAMELS-IllustrisTNG, the CAMELS-SIMBA XSB profile is dependent on the SN energy. However, increasing $B_{\rm SN1}$ leads to a decrease in XSB, mainly in the outer CGM. Increasing the mass loading factor increases the amount of gas being pushed out by the galactic wind. This is contrary to the $A_{\rm SN1}$ dependence in CAMELS-IllustrisTNG, where $A_{\rm SN1}$ represents the wind energy instead of the wind mass outflow.  With other parameters fixed at the fiducial value of 1, varying $B_{\rm SN1}$ alone systematically underpredicts the eRASS observations, especially in the lowest stellar mass bin. 

The second row in Figure~\ref{fig:simba_xsb_profiles} shows the comparison of the eRASS CGM XSB profiles with the SIMBA profiles of varying $B_{{\rm SN2}}$, which represents the speed of the SN wind, similar to $A_{{\rm SN2}}$ in CAMELS-IllustrisTNG. The SIMBA XSB profiles are most sensitive to $B_{{\rm SN2}}$. In contrast to $B_{{\rm SN1}}$, increasing $B_{{\rm SN2}}$, i.e. the SN wind speed, leads to a higher XSB. This is likely because increasing the SIMBA SN wind inhibits the growth of SMBH, leading to suppressed AGN feedback and thus to the retention of more CGM gas \citep{medlock24b}. To match the eRASS XSB profiles, the SIMBA wind speed needs to be higher than the fiducial model value. 

The third row in Figure~\ref{fig:simba_xsb_profiles} shows the comparisons of the eRASS CGM profiles with the emulated XSB profiles with varying SIMBA's $B_{\rm AGN1}$, the feedback energy of the AGN. Similarly to the CAMELS-IllustrisTNG, the XSB profile is not very sensitive to the AGN feedback energy. Overall, varying $B_{\rm AGN1}$ alone does not lead to any agreement between the SIMBA and the eRASS XSB profiles.

The fourth row of Figure~\ref{fig:simba_xsb_profiles} shows the comparison of the eRASS CGM XSB profiles with the SIMBA profiles with varying $B_{{\rm AGN2}}$, which represents the speed of the AGN jet. Increasing $B_{{\rm AGN2}}$ results in a decrease in XSB in the inner CGM and a slight increase in the outer CGM for all stellar mass bins. This means that the jet is more effective in lowering the gas density in the inner regions while transferring some gas mass to the outer regions of halos.

\subsubsection{CAMELS-Astrid}\label{sec:astrid}

Figure~\ref{fig:astrid_xsb_profiles} shows the comparisons of the eRASS CGM profiles with the XSB profiles emulated based on the CAMELS-Astrid runs. 

The top row shows the XSB dependence on $C_{\rm SN1}$, the energy of the supernova feedback per unit of SFR. The Astrid XSB profile is more sensitive to $C_{\rm SN1}$. Increasing $C_{\rm SN1}$ leads to a large increase in the normalization of XSB, qualitatively similar to CAMELS-Illustris. With other parameters fixed at the fiducial value of 1, varying $C_{\rm SN1}$ alone always overpredicts the eRASS observation for all stellar mass bins. 

The second row shows the comparison of the eRASS CGM profile with the CAMELS-Astrid profile with varying $C_{\rm SN2}$, which represents the SN wind speed. Compared to CAMELS-IllustrisTNG and CAMELS-SIMBA, the XSB profiles in CAMELS-Astrid are very sensitive to SN wind speed. Increasing the $C_{\rm SN2}$ parameter increases the XSB, suggesting that increasing the SN wind speed suppresses the AGN feedback, leading to a greater retention of the CGM gas.  

The third row shows the comparison of the eRASS CGM profile with the CAMELS-Astrid profile with varying $C_{\rm AGN1}$, which represents the kinetic AGN feedback energy. Compared to CAMELS-IllustrisTNG and CAMELS-SIMBA, The XSB profiles in CAMELS-Astrid is quite sensitive to AGN feedback energy. Increasing the $C_{\rm AGN1}$ parameter decreases XSB, suggesting that the kinetic AGN feedback is effective in removing the hot gas from the halo. Similarly as $C_{\rm SN1}$, varying $C_{\rm AGN1}$ alone always overpredicts the eRASS observation for all stellar mass bins. 

The fourth row shows the comparison of the eRASS CGM profile with the CAMELS-Astrid profile with varying $C_{\rm AGN2}$, representing the energy of the thermal AGN feedback. Increasing $C_{\rm AGN2}$ leads to higher XSB, but the trend is not monotonic. As $C_{\rm AGN2}$ increases to about 2, the trend reverses, where increasing $C_{\rm AGN2}$ leads to a decrease in XSB normalization. This is because increasing the thermal AGN feedback increases the thermal energy of the gas, leading to adiabatic expansion of gas, thereby reducing the overall gas density and the XSB. 

\begin{figure*}[htbp]
    \centering
    \includegraphics[width=1.0\textwidth]{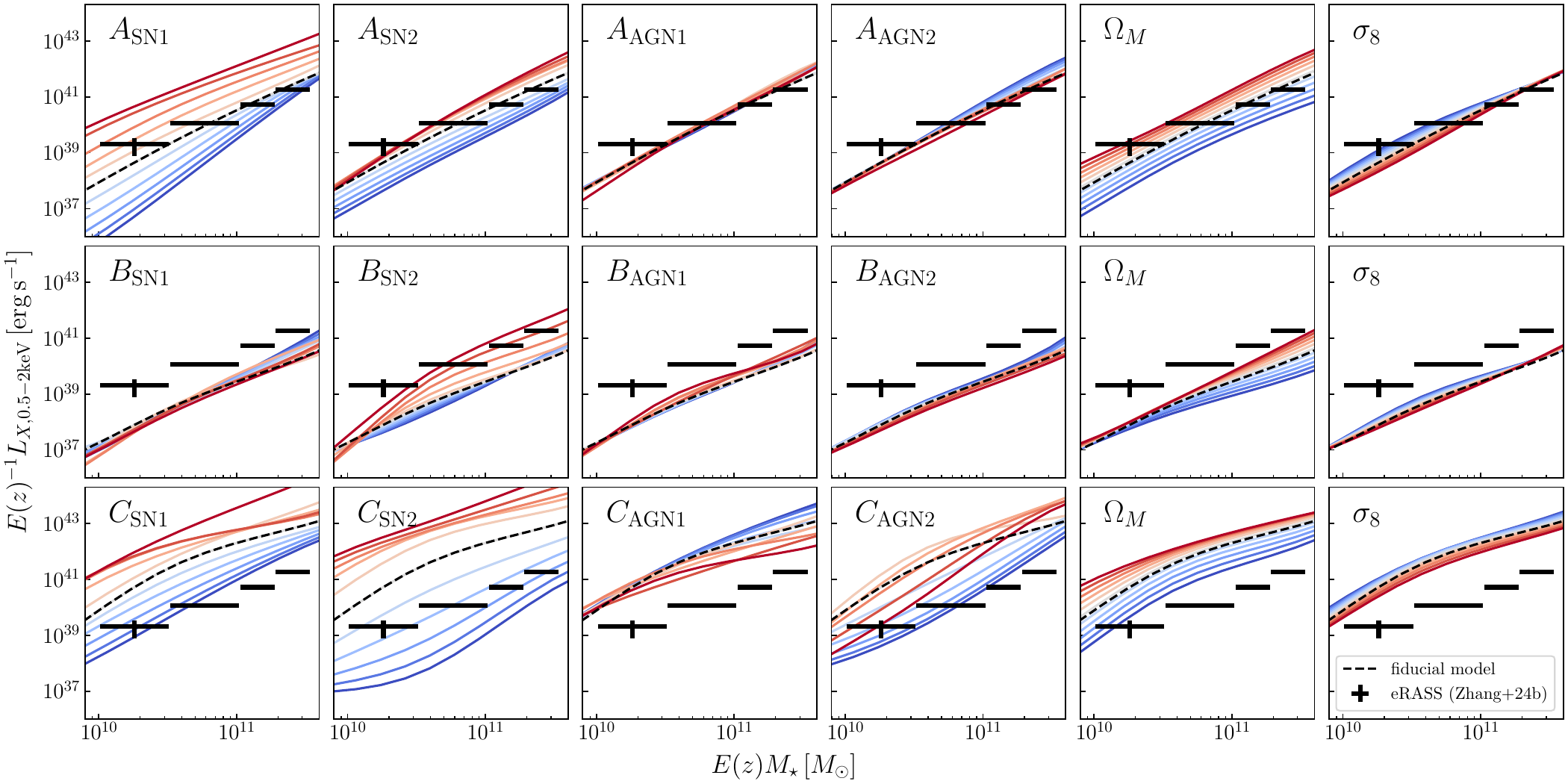}
    \caption{Comparison of the $L_X-M_\star$ relation. The black data points are the eRASS data with 1$\sigma$ errorbars \citep{eRASS_CGM2}. The colored lines are the relations for the CAMELS-IllustrisTNG (top), CAMELS-SIMBA (middle), and CAMELS-Astrid (bottom) with varying values for the 4 feedback parameters and $\Omega_M$ and $\sigma_8$, showed in panels from left to right. The color indicate the strengths of the feedback parameters as in Figures~\ref{fig:tng_xsb_profiles},\ref{fig:simba_xsb_profiles}, and \ref{fig:astrid_xsb_profiles}, with red (blue) represents the max (min) feedback values. The black dashed line indicate the fiducial relation with all feedback strength set to unity.  }
    \label{fig:Lx_Mstar}
\end{figure*}

\subsection{X-ray Luminosity--Stellar Mass Relation}

Figure~\ref{fig:Lx_Mstar} shows the comparison of $L_X-M_\star$ between CAMELS-IllustrisTNG, CAMELS-SIMBA, CAMELS-Astrid and eRASS \citep{eRASS_CGM2}. We see a similar picture in the dependence on the 4 feedback parameters as in the XSB profiles. For CAMELS-IllustrisTNG, the relation is most sensitive to $A_{\rm SN1}$, followed by $A_{\rm SN2}$, $A_{\rm AGN2}$, and $A_{\rm AGN1}$. The fiducial TNG model underpredicts the eRASS data at low stellar mass but overpredicts at high stellar mass, but all within $2\sigma$ of the observation. Changing $A_{\rm SN1}$ leads to a larger change in the slope of the scaling relation, since X-ray luminosities in low stellar mass halos are more sensitive to SN feedback than higher stellar mass halos. 

For CAMELS-SIMBA, the $L_X-M_\star$ relation is most sensitive to $B_{\rm SN1}$ and $B_{\rm SN2}$, however, their trends are reversed. Higher $B_{\rm SN1}$ leads to lower $L_X$, but higher $B_{\rm SN2}$ leads to higher $L_X$. The impact of changing $B_{\rm SN1}$ is stronger in high stellar mass, thus varying $B_{\rm SN1}$ also changes the slope in the relation. Changing $B_{\rm SN2}$ leads to almost uniform change in $L_X$ across the stellar mass bin, except at the lowest stellar mass, where $L_X$ is not particularly sensitive to $B_{\rm SN2}$. The fiducial CAMELS-SIMBA model underpredicts the eRASS data at all stellar masses. 

CAMELS-Astrid shows a stronger sensitivity to feedback parameters in the $L_X-M_\star$ relation, compared to CAMELS-TNG and CAMELS-SIMBA. The relation is most sensitive to $C_{\rm SN2}$, followed by $C_{\rm SN1}$, $C_{\rm AGN2}$, and $C_{\rm AGN1}$. The fiducial CAMELS-Astrid model overpredicts the eRASS $L_X$ over all stellar masses. Increasing the $C_{\rm SN1}$, $C_{\rm SN2}$, and$C_{\rm AGN1}$ feedback parameters leads to a slight decrease in the slope of the scaling relation, as the increases in feedback increase the X-ray luminosities in lower stellar mass halos more due to their shallower potential wells. While increasing $C_{\rm AGN2}$ leads to higher normalization in $L_X-M_\star$, for $C_{\rm AGN2} \geq 1.5$, the normalization decreases as strong thermal AGN feedback decreases the gas density and thus XSB in the inner CGM (see fourth row in Figure~\ref{fig:astrid_xsb_profiles}). 

\subsection{Dependence on $\Omega_M$ and $\sigma_8$}

For all three simulation suites, we see similar dependencies of the XSB profile and the $L_X-M_\star$ relation to $\Omega_M$ and $\sigma_8$. Increasing $\Omega_M$ leads to higher XSB and $L_X$ values, due to the increase in the virial temperature of the CGM, as the increase in $\Omega_M$ leads to higher halo masses for a given stellar mass. Here, the $\Omega_M$ dependence is degenerate with the feedback parameters, especially with the SN parameters across the suites. In contrast, both XSB and $L_X$ are not sensitive to $\sigma_8$. 

\subsection{Constraints on Feedback Parameters}\label{sec:mcmc}

\begin{figure}
    \centering
    \includegraphics[width=0.5\textwidth]{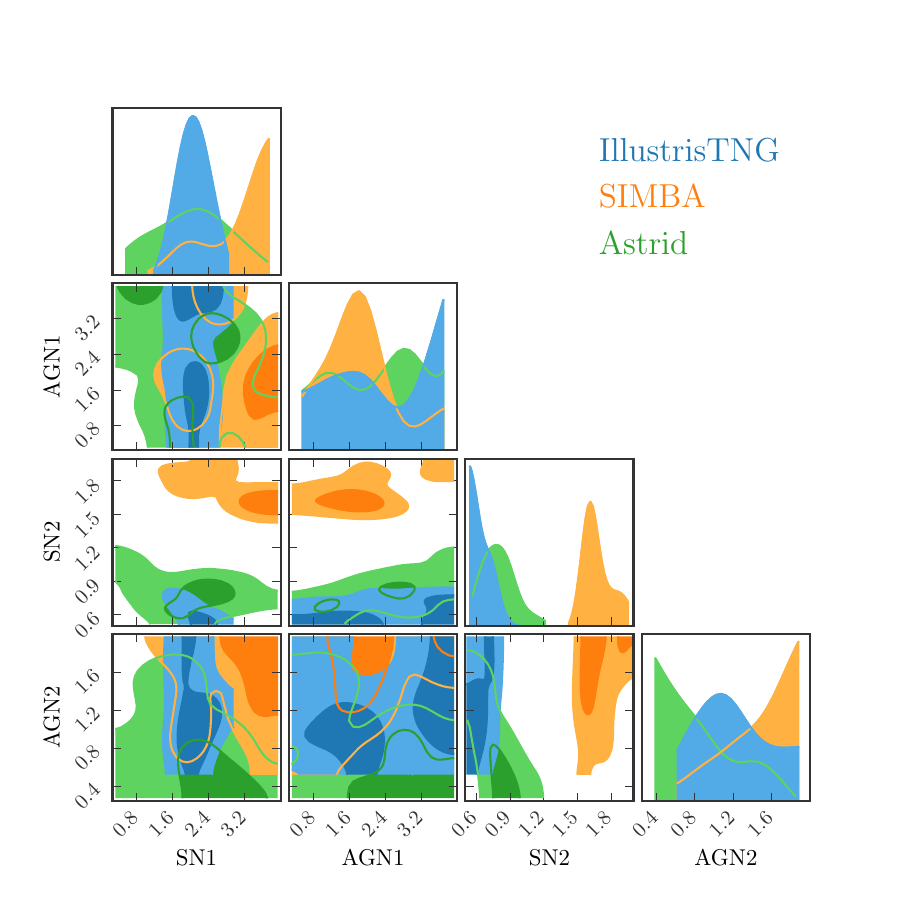}
    \caption{Posteriors of the best-fit parameters from MCMC for CAMELS-IllustrisTNG (blue), CAMELS-SIMBA (orange), and CAMELS-Astrid (green). The contours represent $1$ and $2\sigma$ of the uncertianties in the parameters. }
    \label{fig:mcmc_constraints}
\end{figure}

We apply the emulator to the eRASS data set to obtain constraints on the feedback parameters. We follow the conventional Monte Carlo Markov Chain (MCMC) to infer the posterior distributions of the parameters given the eRASS data. Inference is performed by maximizing a Gaussian likelihood function. Specifically, the likelihood function is

\begin{equation}
\ln \mathcal{L}(L_X|\mathbf{\theta}) \propto - \sum_i \frac{\left(L_{X,i}^{\rm mod}(M_\star, z;\mathbf{\theta})-L_{X,i}(M_\star, z\right)^2}{2\sigma_i^2},
\end{equation} for the $L_X-M_\star$ relation
where $L_X^{\rm mod}$ is the emulated X-ray luminosity, with parameters $\mathbf{\theta} = \left(A_{\rm SN1}, A_{\rm SN2}, A_{\rm AGN1}, A_{\rm AGN2}\right)$, $L_{X,i}$ is the eRASS data with error $\sigma_i$ at the $i$-th stellar mass bin.  Because there are only 4 data points in the eRASS data, we do not fit for $\Omega_M$ and $\sigma_8$. Including them would result in 6 independent parameters, requiring 2 more data points. We fixed $\Omega_M = 0.3089$ and $\sigma_8=0.8102$ \citep{planck2018}.  
The uniform prior for each feedback parameter is chosen to match the range explored in the simulations. We run MCMC for $10^5$ steps. For each chain, we discard the first $5\times 10^4$ steps to make sure that the posteriors are not affected by the initial values of the parameters.  We determined that the chains converged by making sure that the means and variances of the parameter values in the last 5000 steps remain essentially unchanged.

\begin{table}
\caption{Best-fit Feedback Parameters}
\begin{tabular}{|c | c | c | c |}
 \hline
 \textbf{Parameter} & \textbf{IllustrisTNG} & \textbf{SIMBA} & \textbf{Astrid} \\
 \hline
SN1 & $2.07^{+0.35}_{-0.34}$ & $3.45^{+0.35}_{-0.38}$ & $2.07^{+0.82}_{-1.00}$  \\

SN2 & $0.6^{+0.13}_{-0.07}$ & $1.63^{+0.20}_{-0.05}$ & $0.78^{+0.13}_{-0.13}$ \\

AGN1 & $2.18^{+1.56}_{-1.14}$ & $1.79^{+0.55}_{-0.80}$ & $2.42^{+0.84}_{-1.39}$ \\

AGN2 & $1.15^{+0.53}_{-0.54}$ & $1.62^{+0.27}_{-0.53}$ & $0.67^{0.65}_{-0.30}$ \\
\hline
\end{tabular}\label{tab:bf}
\end{table}

\begin{figure*}
    \centering
    \includegraphics[width=0.45\textwidth]{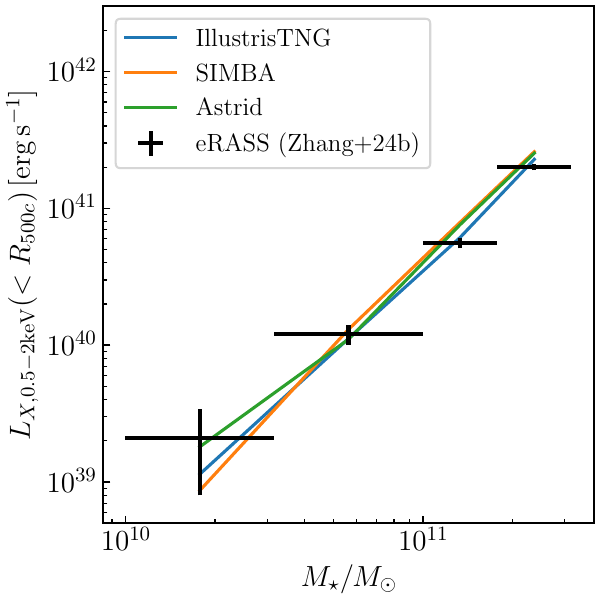}
    \includegraphics[width=0.45\textwidth]{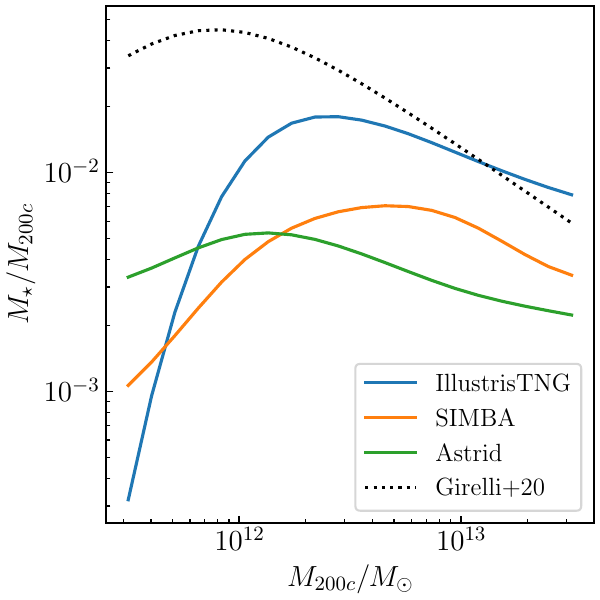}
    \caption{
    Left panel: The best-fit $L_X-M_\star$ scaling relations for CAMELS-IllustrisTNG (blue), CAMELS-SIMBA (orange), and CAMELS-Astrid (green), obtained by fitting against the eRASS data (black data points) from \citet{eRASS_CGM2}. 
    Right panel: the corresponding stellar mass fraction as a function of halo mass for CAMELS-IllustrisTNG (blue), CAMELS-SIMBA (orange), and CAMELS-Astrid (green) compared to the empirical fit to observed stellar mass fraction from \citet{girelli_etal20} (dashed line). 
    }
    \label{fig:best_fit}
\end{figure*}

Figure~\ref{fig:mcmc_constraints} shows the posterior distributions of the feedback parameters for CAMELS-IllustrisTNG, CAMELS-SIMBA, and CAMELS-Astrid, respectively. The eRASS data provide reasonably good constraints on the feedback parameters. The reduced $\chi^2$ values for the best-fit models are $0.83, 4.00$, and $3.01$ for CAMELS-IllustrisTNG, CAMELS-SIMBA, and CAMELS-Astrid respectively, compared to the those for the fiducial models: $4.17, 313$, and $666$. 
Figure~\ref{fig:best_fit} shows the eRASS best-fit scaling relations for CAMELS-IllustrisTNG, CAMELS-SIMBA, and CAMELS-Astrid, compared to the eRASS measurements from \citet{eRASS_CGM2}, and the best-fit values for the feedback parameters and their uncertainties are summarized in Table~\ref{tab:bf}.

For CAMELS-IllustrisTNG, the eRASS data prefer slightly stronger SN1 (SN energy per SFR) and slightly weaker SN2 (stellar wind speed) feedback than the fiducial IllustrisTNG model. There is no constraint on AGN1 (kinetic AGN feedback energy) and AGN2 (AGN feedback frequency) is almost the same as the fiducial value. 
For CAMELS-SIMBA, the model requires stronger SN and AGN feedback than the fiducial run to match the eRASS X-ray observations. 
On the other hand, for Astrid, the eRASS data prefer higher SN1 and AGN1 feedback, i.e., higher stellar feedback energy, and lower kinetic AGN feedback energy, respectively, but prefer lower values for SN2 and AGN2, representing SN wind speed and thermal AGN feedback energy, respectively. 

\subsection{Implications for the Stellar Mass--Halo Mass relation}

The right panel in Figure~\ref{fig:best_fit} shows the stellar mass-halo mass relations for CAMELS-IllustrisTNG, CAMELS-SIMBA, and CAMELS-Astrid using their respective best-fit feedback parameters from the MCMC analysis performed on the eRASS data in Section~\ref{sec:mcmc} (solid curves). 
We compare them to the empirical fit to the observed stellar mass-halo mass relation \citep{girelli_etal20}. There are differences between the best-fit CAMELS-eRASS relations with the observation. 
For CAMELS-IllustrisTNG, the eRASS best-fit stellar mass-halo mass relation matches with the fiducial one for halo masses $\log_{10}(M_{200c}/M_\odot) > 13$, below this mass scale, the predicted stellar mass fraction becomes smaller than the observed value. For both CAMELS-SIMBA and CAMELS-Astrid, the predicted stellar mass fraction is smaller than observations. 

The disagreement between the eRASS-best-fit stellar mass fraction and the observed stellar-mass halo mass relation, suggests that the feedback models in the CAMELS simulations are not able to simultaneously match both the eRASS X-ray CGM and observed stellar properties. There are also disagreements in the properties of the CGM between cosmological simulations and recent thermal and kinetic SZ observations of the CGM with the Atacama Cosmology Telescope \citep[ACT,][]{amodeo_etal21}, which prefers higher thermal pressure gas density and thus stronger feedback \citep{moser_etal22}. Although for ACT, the differences showed up at larger halo-centric radii ($r\gtrsim 1$~Mpc) and at higher halo masses ($M_{200c}\gtrsim 10^{13} M_\odot$). 

This apparent tension between simulations and observations can be due to (1) systematics in the stacked eRASS observations; or (2) inadequacies of the feedback models in the cosmological simulations.

For (1), the eRASS X-ray CGM emission can potentially be overestimated. Specifically, X-ray binaries in the interstellar medium can contaminate the CGM due to the relatively wide point spread function ($\sim 27 \arcsec$) of eROSITA/SRG. Incomplete removal of background X-ray emissions from nearby galaxy groups can also boost the CGM signal. Since the observed eRASS XSB profiles and X-ray luminosities are derived from the stacking of spectropscopically selected galaxies, they are also susceptible to uncertainties in the stacking procedure. The stacked values can be biased by a few outliers in the stacking sample; the central galaxies selected for stacking can also be misclassified. For example, the X-ray properties of the CGM from the eROSITA eFEDS measurements have been shown to depend on the star formation rate (SFR) and the morphology of its host galaxy \citep{chadayammuri_etal22,comparat_etal22}. Specifically, star-forming galaxies are shown to be more X-ray bright than quiescent galaxies. Preferentially selection of star-forming galaxies in the stacking procedure can lead to overestimates in the X-ray emissions. However, recent eRASS measurements \citep{eRASS_CGM3} show that for Milky Way mass systems, star-forming and quiescent galaxies have similar X-ray luminosities, while for more massive galaxies ($M_\star > 10^{11} M_\odot$), the X-ray luminosities of quiescent galaxies surpass those of star-forming systems. These potential systematics can be addressed with observations of the CGM of individual galaxies with XMM-{\em Newton} and {\em Chandra} that have higher angular resolution. SZ observations, which do not suffer from contamination by X-ray binaries, will also be useful for assessing these systematics. 
If the X-ray CGM is indeed overestimated, accounting for the contaminants will alleviate the need for stronger feedback, thus reducing the tension in the stellar mass fractions between simulations and the one inferred from eRASS. 

For (2), as the CAMELS simulations have lower resolution compared to their corresponding flagship simulations (TNG100, \citealt{pillepich_etal18}, SIMBA, \citealt{dave19}, Astrid, \citealt{astrid}), the CAMELS versions show slightly lower X-ray luminosity than their flagship counterparts (compare the dashed lines in Figure~\ref{fig:Lx_Mstar} with Figure~2 in \citealt{eRASS_CGM2}), thus the level feedback may not be required to be as high as the CAMELS version for these flagship simulations, potentially reducing the tension with eRASS. 
In addition, the X-ray CGM properties can also depend on other parameters in the subgrid models that have not been explored in the current paper. We will address this issue with CAMELS simulations with more varied parameters, such as the 1P28 \citep{medlock24b} and the SB28 sets \citep{camels_astrid}.  

%Finally, the stellar mass - halo mass relation in Figure 6 (which is an excellent inclusion) warrants additional discussion: there are tensions between the observables that pull in different physical directions and make this a hard problem to solve! To my eye, it appears that the stronger stellar feedback required to suppress the AGN feedback also suppresses stellar mass growth.

\section{Conclusions}

In this paper, we use the CAMELS simulation suite to construct emulators of the X-ray surface brightness (XSB) profile and the X-ray luminosity-stellar mass ($L_X-M_\star$) relation at the massive galaxy scale (with $\log_{10} M_\star/M_\odot \in [10, 11.5]$) to model the dependence of the circumgalactic medium (CGM) on feedback due to supernovae and active galactic nuclei in subgrid models from three modern cosmological hydrodynamical simulations: IllustrisTNG, SIMBA, and Astrid. We compare the emulated XSB profiles and the $L_X-M_\star$ relation to the recent stacked measurements from the eROSITA All Sky Survey \citep[eRASS,][]{eRASS_CGM1, eRASS_CGM2}. Here are our findings:
\begin{itemize}
    \item The XSB and $L_X$ of all 3 models explored depend more sensitively on SN feedback than AGN.
    In particular, the changes in the efficiency of AGN feedback have a lower impact on the X-ray observables.
    
    \item The variations in the predicted X-ray properties among the 3 fiducial models are significant. For CAMELS-IllustrisTNG, the X-ray properties of the CGM match the eRASS observation best at higher stellar mass (M31-type galaxies and more massive ones). CAMELS-SIMBA underestimates the eRASS XSB profiles and $L_X$ at all stellar masses by an order of magnitude, and CAMELS-Astrid overpredicts the eRASS results by an order of magnitude. 
    
    \item We derive the best-fit values of the feedback parameters of the three models using the $L_X-M_\star$ measurements from eRASS. While the eRASS data prefers stronger stellar feedback than the fiducial models in the CAMELS simulations, the stellar mass-halo mass relations using the best-fit values do not match with observed stellar mass-halo mass relation. 

\end{itemize}

Although it seems counter-intuitive that the X-ray emissions are more dependent on SN and AGN feedback for CAMELS-IllustrisTNG and CAMELS-Astrid, in detail SN feedback is mainly responsible for regulating the AGN feedback that quenches star formation as SN feedback limits the amount of gas available for accreting onto the AGN \citep{lee_etal24, medlock24b}. This effect turns out to be more important than varying the amount of AGN feedback energy. To a lesser extent, more SN feedback also increases the metallicity of CGM, thereby increasing the X-ray emission. In addition, the small box size of the CAMELS simulation also limits the number of more massive halos ($\sim 10^{13}M_\odot$) that are more susceptible to AGN feedback. Upcoming CAMELS simulations with larger box sizes, or zoom-in simulations \citep[e.g.,][]{lee_etal24} have larger numbers of more massive galaxies, groups, and clusters, and will provide tighter constraints on AGN feedback physics.

Although modern cosmological hydrodynamical simulations often calibrate their galaxy formation model using observation data of galaxies, such as the stellar mass function and the stellar mass-halo mass relation, the equally important CGM properties have not been sufficiently constrained or compared to observations. Our work demonstrates that X-ray observables are powerful in inferring and constraining the baryon cycles in massive galaxies. As next steps, in addition to the XSB profiles and X-ray luminosities studied in this work, temperature profiles, metallicity profiles, and spatial and kinematic structures will provide even more powerful constraints on baryonic feedback \citep{schellenberger_etal24, zuhone_etal24}. SZ and Radio observations will also provide complementary probes of the CGM \citep[e.g.,][]{medlock24a,singh_etal24}.  Forward modeling with light-cone simulations \citep[][]{shreeram_etal24,lau_etal24} will become useful for accounting for observational systematics in the multiwavelength CGM observations. Combined with optical properties and SZ measurements, these X-ray constraints will serve as important input for refining feedback models in cosmological simulations and improving our understanding of galaxy formation.

\section*{acknowledgments}
We thank the referee for useful comments and feedback. 
We also thank Yi Zhang, Johan Comparat, and the rest of the eROSITA Collaboration for providing us with the eRASS data and for stimulating discussions. DN and BDO acknowledge the support of the NSF grant AST-2206055. \'A.B. acknowledges support from the Smithsonian Institution and the Chandra Project through NASA contract NAS8-03060. 
DAA acknowledges support by NSF grant AST-2108944, NASA grant ATP23-0156, STScI grants JWST-GO-01712.009-A and JWST-AR-04357.001-A, Simons Foundation Award CCA-1018464, and Cottrell Scholar Award CS-CSA-2023-028 by the Research Corporation for Science Advancement. The Flatiron Institute is supported by the Simons Foundation.

\software{
emcee \citep{emcee}
}

\bibliography{references}{}
\bibliographystyle{aasjournal}

\end{document}